*Original Article*

# ASCNet-ECG: Deep Autoencoder based Attention aware Skip Connection network for ECG filtering

Raghavendra Badiger[1], M. Prabhakar[2]

[1,2]*Department of CS & E, REVA University, Bengaluru, Karnataka, India.*

[1]*Corresponding Author : raghavendra.badiger1@gmail.com*

***Abstract*** *- Currently, the telehealth monitoring field has gained huge attention due to its noteworthy use in day-to-day life. This advancement has led to an increase in the data collection of electrophysiological signals. Due to this advancement, electrocardiogram (ECG) signal monitoring has become a leading task in the medical field. ECG plays an important role in the medical field by analysing cardiac physiology and abnormalities. However, these signals are affected due to numerous varieties of noises, such as electrode motion, baseline wander and white noise etc., which affects the diagnosis accuracy. Therefore, filtering ECG signals became an important task. Currently, deep learning schemes are widely employed in signal-filtering tasks due to their efficient architecture of feature learning. This work presents a deep learning-based scheme for ECG signal filtering, which is based on the deep autoencoder module. According to this scheme, the data is processed through the encoder and decoder layer to reconstruct by eliminating noises. The proposed deep learning architecture uses a modified ReLU function to improve the learning of attributes because standard ReLU cannot adapt to huge variations. Further, a skip connection is also incorporated in the proposed architecture, which retains the key feature of the encoder layer while mapping these features to the decoder layer. Similarly, an attention model is also included, which performs channel and spatial attention, which generates the robust map by using channel and average pooling operations, resulting in improving the learning performance. The proposed approach is tested on a publicly available MIT-BIH dataset where different types of noise, such as electrode motion, baseline water and motion artifacts, are added to the original signal at varied SNR levels. The outcome of the proposed ASCNet is measured in terms of RMSE and SNR.*

## 1. Introduction

In a recent study, World Health Organization (WHO) estimated that non-communicable diseases (NCDs) account for 41 million deaths annually and 71% of all fatalities worldwide. The high impact is noticed in individuals in the age range of 30-69 years. The report has submitted that 15 million people die prematurely every year due to NCDs. Cardiovascular diseases in NCDs cause the majority of mortalities. In 2022, WHO reported 17.9 million mortalities due to NCDs, 9 million due to cancer, 3.9 million due to respiratory illness, and 1.6 million due to diabetes [1]. These studies have reported that cardiovascular diseases (CVDs) are the foremost mortality source worldwide. Recently, American Health Association presented a report on cardiovascular diseases and estimated 23 million mortalities by 2030 due to cardiac abnormalities [2,3].

Currently, these abnormalities are affecting developed and developing countries. This trend of mortalities is observed in low and middle-income nations, which accounts for 85% of fatalities of worldwide cases. Its effect is increasing in urban lifestyles due to several factors such as busy work schedules, stress, imbalanced diet, smoking habit, alcohol etc. These habits are considered the serious cause of several heart-related diseases such as blood pressure (BP), cardiac arrest, hypertension, heart attacks etc. [4].

Thus, early detection and diagnosis can help to reduce the mortality rate. Therefore, monitoring heart health is a primary task to identify the abnormality in the heart.

Electrocardiogram (ECG) is a widely adopted technique to record and measure the heart's electrical activities. Moreover, current developments in biomedical health monitoring have gained huge attention due to the popularisation of wearable devices to monitor health. These wearable technologies are widely adopted and mainly based on ECG processing techniques. Several devices have been developed for ECG analysis such as Zhang et al. [5] used a wearable device for fetal ECG monitoring, Wu et al. [6] introduced IoT based ECG monitoring system and integrated on a T-shirt, and Beach et al. [7] developed wrist-worn ECG monitoring system based on IoT systems. Thus, ECG analysis plays an important role.

### 1.1. ECG Signal Generation from Heart Activity and its Morphology

It is a technique used for recording the strength and timing of the heart's electrical activities. Electrocardiography generates the graph of voltage versus time of the heart's electrical activities using electrodes placed on the skin. Below, figure 1 depicts the heart structure along with the various nodes that generate the signals, and figure 2 depicts the complete ECG signal.

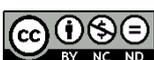

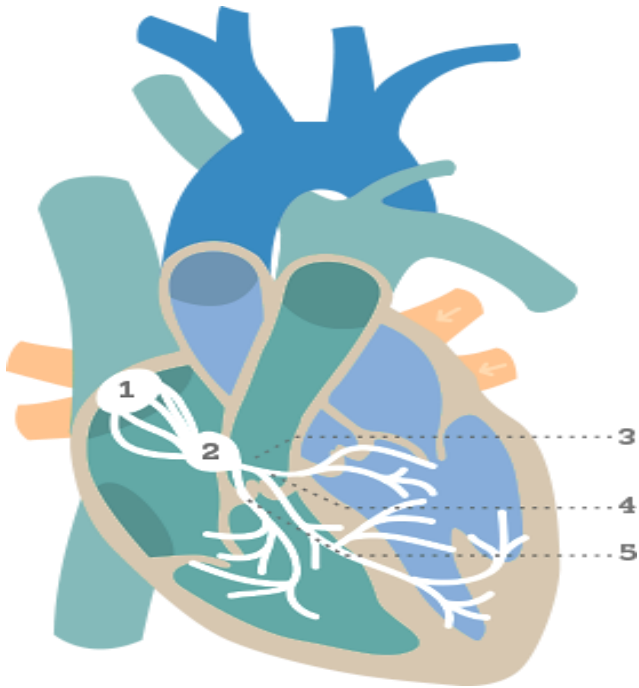

**Fig. 1 Human heart structure**

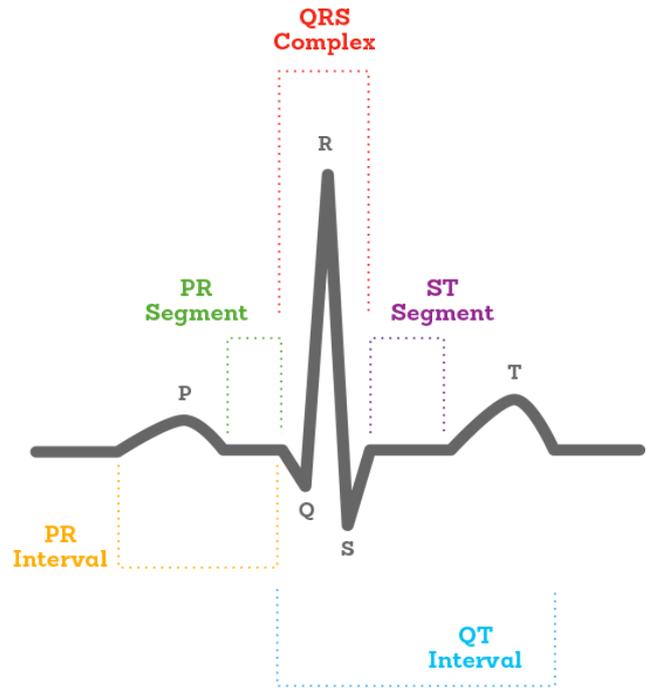

**Fig. 2 ECG signal**

The complete process can be described as follows:

- Point (1), depicted in figure 1, denotes the sinoatrial node. It is positioned in the right atrium.
- This node is connected to the left atria and is responsible for contracting and pumping blood into the ventricles.
- The signal generated from this movement of blood is recorded as P-wave.
- The generated signal moves from the atria to the ventricles through node (2), called as atrioventricular (AV) node. During this process, the signal slows down until the ventricles are filled with blood. This slowing-down process is recorded as a flat line at the end of the P wave and starting of the Q wave.
- The PR interval denotes the time in seconds from the P wave to the start of the QRS complex. The PR segment denotes the electrical signal conduction through the atria.
- After leaving this AV node, the signal follows a pathway called the bundle of His (3), right and left bundle branches (4) and (5), respectively.
- The complete movement of signal across heart ventricles causes the contraction resulting in blood pumping to the lungs and body. This signal is recognised as a QRS wave.
- Once the QRS complex is completed, the ventricles acquire their normal state, denoted as T wave.
- Muscle relaxation results in stopping the contraction and allows the atria to be filled with blood. This process is repeated in each heartbeat.
- The ST signal connects the QRS and T wave, which shows the electrical recovery of the ventricles.
- The QT interval shows the duration in which ventricles are simulated and recovered after the complete simulation.

*1.2. Current research and challenges in ECG signal processing*

Due to their wide use in the biomedical field, several researchers have carried out to analyse heart activities effectively. ECG signals are widely adopted in various applications such as Caesarendra et al. [10] developed CNN based model to detect normal heart health. Supraventricular arrhythmia using ECG signal, Parmar et al. [11] implemented the Fourier decomposition method and modulated filter bank to detect hypertension, Ayashm et al. [12] analysed ECG signal[8] to detect sleep Apnea, Ramkumar et al. [13] used ECG processing for arrhythmia classification. However, the amplitude of ECG signals is low as 0.1mV to 2.5mV thus, it is vulnerable to environmental noise, which degrades the signal quality [9,27]. Moreover, during signal acquisition, the original signal gets contaminated due to numerous varieties of noises, such as:

- Power line interference: it is a type of noise component of 50 Hz/60 Hz depending upon the frequency of the power supply.
- Motion artifacts: this is the noise which is generated by the patient's movement during signal acquisition. This affects the impedance between the electrode and the skin. This noise causes a 100-500ms long delay.
- Electrode-contact noise: it is generated due to inappropriate contact between the body and electrodes with ~1Hz frequency.
- Muscle contraction: generally, the muscle contraction produces noise with an amplitude of 10% of regular peak-to-peak ECG signal and frequency up to 10kHz. The duration of this signal is 50 ms.
- Baseline wanders: this noise is induced due to respiration activity, having a frequency up to 0.5Hz, and the amplitude of this noise is almost ~15% of the ECG amplitude.

Table 1. Common ECG artifacts with description, causes and example

| Artifacts | Description | Cause of Artifact | Example |
|---|---|---|---|
| (1) Wandering Baseline | A slow wander of the baseline | (i) Body movement<br>(ii) Respiratory swing | 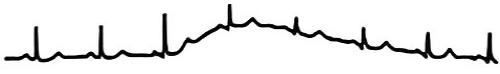 |
| (2) AC Interference | Varying amplitude of ECG and indistinct isoelectric baseline | (i) Elect. Power leakage<br>(ii) Improper equipment grounding<br>(iii) Close proximity of other electrical equipment | 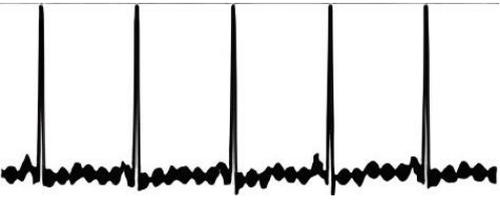 |
| (3) Muscle Tremor | Narrow and rapid spike of ECG | (i) Effect of EMG signal<br>(ii) Shivering<br>(iii) Parkinson's disease | 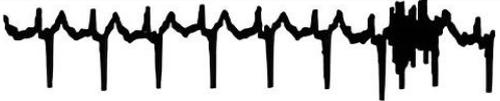 |
| (4) Motion Artifact | Large swing in the baseline, uncertainty of large amplitude signals | (i) Effect of epidermal signal<br>(ii) Stretching the epidermis<br>(iii) Coughing<br>(iv) Ambulation | 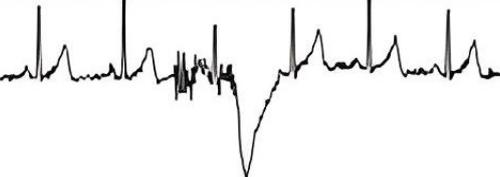 |

Some of the commonly occurring artifacts in ECG signals are demonstrated below the given table, along with its cause [28].

Due to this issue, ECG signal processing and removing the different kinds of artifacts has become a challenging task and the most important task to improve the diagnosis. Moreover, denoised or filtered signal helps to accurately identify the peaks of the ECG signal, which is useful in heart abnormality detection. Several techniques have been introduced in this field for ECG peak detection and filtering. Some of the recent techniques and their drawbacks and challenges are described in section II.

*1.3. Problem Definition*
The demand for ECG-based health monitoring systems has increased drastically. However, data acquisition is a crucial stage where signal quality is distorted due to several types of noises, as mentioned before. Therefore, removing the noise can be beneficial to facilitate differential diagnosis. Furthermore, peak detection also plays an important role in analysing heart activity. For example, Warmerdam et al. [21] used ECG peak detection for heart rate analysis, and Bae et al. [22] used ECG analysis for arrhythmia analysis. However, the peak detection performance is influenced by noise.

*1.4. Contribution and Novelty of the Work*
The main objective of this investigation is to articulate a novel automated approach for ECG signal filtering which can improve peak detection. The main contribution of this research is as follows:

- To develop a deep learning-based architecture for ECG denoising based on a Deep Autoencoder scheme where data is processed through the encoder and decoder phases.

- This model uses a modified ReLU function to handle the noise variations. According to this model, the input is divided into two parts and combined together by applying average pooling.
- This model presents a new skip connection model that connects the encoder's outputs to the decoder.
- Finally, an attention mechanism is presented, which uses channel and spatial attention mechanisms.

The traditional methods focus only on a single type of domain, i.e., time or frequency domain, which affects the learning process. Therefore, a combined model is introduced, which considers time and frequency domain problem formulation. Further, the training performance of the deep learning model is enhanced by incorporating an improved Leaky ReLU function. Furthermore, an attention-based mechanism is also introduced, which considers the channel and spatial information to exploit the inter and intra-channel attributes

*1.5. Article Organisation*
The rest of the manuscript is arranged in the following sections: Section II presents a brief literature review of existing techniques, section III presents the proposed solution, section IV describes the outcome of the proposed approach and its comparative analysis, and finally, section V presents the presents the concluding remarks.

## 2. Literature Survey
This segment briefly discusses recent procedures to filter the ECG signal. Generally, the ECG classification methods face this problem, affecting the accuracy of ECG classification [15]. This scenario occurs because ECG signals are weak in amplitude and easily get contaminated due to other sources. Currently, several techniques are present such as the Butterworth filter[14], adaptive filters [29], Wiener filter, Kalman filter etc. the adaptive filters are considered self-designing filters which adopt the Least Mean Square method to adjust the weights of the filter to minimise the error. Weiner filter requires a statistical noise component to perform filtering; the Kalman filter applies state prediction and implements an observation model to analyse the filtering. These filters are widely adopted to remove white noise [18].

Similarly, time domain and frequency domain-based filtering techniques are widely adopted for ECG signal filtering. These techniques use the Fourier transform, which provides only spectral information, leading to a loss in the information. To overcome this issue, Short Time Fourier Transform (STFT) technique is also utilised, which generates both time and frequency domain analysis with the help of the moving window function [18].

Recently, machine learning-based approaches have gained huge attention in the biomedical field. These techniques are adopted in ECG signal filtering. Rasti-Meymandi et al. [16] discussed the importance of deep learning methods in noise removal and reported that traditional methods consider only 1D time series of ECG signals. Thus, the authors introduced a deep learning-based approach to process 2D signals where ECG cardiac cycles are stacked together to generate a 2D signal processed through the CNN model. Moreover, it uses correlation analysis of cardiac cycles to improve denoising performance. The CNN model uses local and non-local cycle observation to measure the correlation between cardiac cycles. Mourad et al. [17] presented a local filtering approach to remove wideband noise from ECG signals. According to this process, the ECG signal is modelled into different components in the time domain and frequency domain. The segmentation method is applied where it is assumed that each segment consists of one dominant component. This signal is processed through the successive filtering scheme. During reconstruction output of successive ECG is collected using the ideal filter.

Romero et al. [20] reported that baseline wander is the most commonly occurring noise. Therefore, a deep learning-based scheme that uses multipath modules to extract the features is introduced. The multipath modules consist of convolution layers which are placed at the same level. The backpropagation algorithm is also applied to select the path and weights at this stage.

Similarly, peak detection is also widely adopted to analyse heart activities. Warmerdam et al. [21] reported that fetal electrocardiogram (fECG) provides detailed and valuable information about fatal. This method uses the R-peak of the ECG signal. However, R-peak detection is a challenging task. Thus, the authors proposed a multichannel probabilistic method for R-peak detection.

Bae et al. [22] developed an adaptive Medial filtering-based method for R-peak detection for arrhythmias detection. Initially, this model determines the size of the sliding window and median filtering size based on the sampling rate. Later, a median filter is applied in the sliding window.

Zahid et al. [23] took advantage of the machine learning concept and presented a convolutional neural network-based approach for R-peak detection. Authors suggested that traditional methods fail to achieve the desired performance for low-quality signals, leading to false alarms in peak detection. Therefore, the authors presented 1D CNN model along with a verification model to minimise the false alarms.

### 2.1. Research Gap

The previous section has described about several ECG filtering schemes, including time domain and frequency domain filtering methods. The wavelet-based methods provide a feasible solution to this problem, but the appropriate selection remains challenging. Moreover, these algorithms have low efficiency in signal smoothing. Similarly, these techniques are based on the averaging and smoothing operations for converting the signal to frequency to the time domain and time domain to the frequency domain. Currently, deep learning-based schemes are widely adopted where huge training samples are required to achieve the desired performance. Moreover, the complex architecture and high configuration requirement to train deep learning models also pose several challenges.

## 3. Proposed Model

This section presents the proposed solution for ECG filtering by considering different types of noises. Advanced ECG denoising techniques have recently reported the significance of deep learning schemes. Moreover, denoising autoencoders have gained attention in this field of signal filtering.

### 3.1. AE and DAE models

An autoencoder's main aim is to accurately reconstruct the contaminated or corrupted signal while considering the loss function constraints in Deep Learning. Generally, the AE architecture is comprised of two components as an encoder($Enc$) and decoder ($Dec$) modules. The $Enc$ model follows deterministic mapping to map the features of former maps of input feature $x$ to a new representation $y$. Similarly, in the decoder model, latent representation $y$ is mapped to a reconstructed vector $z$. The $y$ and $z$ vectors are represented as:

$$y = \varphi(Wx + b) \\ z = \varphi'(W'y + b') \quad (1)$$

Where $W$ denotes the weight matrix of the encoder and $b$ denotes the bias vector of the encoder, similarly, $W'$ and $b'$ denotes the weight and bias vector of the decoder module. $\varphi$ characterise the activation function for $Enc$ and $\varphi'$ denotes the activation function $Dec$ modules. These parameters are augmented by diminishing the signal reconstruction error. This can be expressed as:

$$L = \arg\min_{\theta} \frac{1}{N} \sum_{i=1}^{N} \|x_i - z_i\|_2^2 \quad (1)$$

Where $N$ denotes the overall sample count, and $\theta$ denotes the parameter set $\{W, b, W', b'\}$

Similarly, DAE is a type of classical AE. In DAE models, the corrupted version of data is provided as input $\tilde{x}$. This signal is constructed with the help of stochastic random probability distribution mapping as $\tilde{x} \sim q(\tilde{x}|x)$. This corrupted signal is mapped with the help of the first part of equation (1) and reconstructed using the second part of the equation (1). In this complete process, these inputs are mapped, the output is generated, and these parameters are trained to minimise the error with the help of the error optimisation function given in equation (2)

### 3.2. Time-frequency Domain Problem Formulation

Initially, the input signal is processed through the STFT scheme to generate the time-frequency domain representation of the ECG signal. Let us consider that a signal is represented as:

$$f(t) = A(t)e^{i2\pi\phi(t)} \quad (2)$$

Where $A(t)$ and $\phi(t)$ denote the amplitude and phase of the ECG signal, respectively. Further, Taylor expansion of signal for $\tau$ close to $t$ can be expressed as:

$$f(\tau) = \exp\left(\sum_{k=0}^{N} \frac{[\log(A)]^{(k)}(t) + i2\pi\phi^{(k)}(t)}{k!}(\tau - t)^k\right) \quad (4)$$

With the help of this, the local frequency estimate $\omega_f(t, \eta)$ can be expressed in the form of Taylor expansion:

$$\omega_f(t, \eta) = \frac{[\log(A)]'(t)}{i2\pi} + \phi'(t)$$

$$+ \sum_{k=2}^{N} \frac{[\log(A)]^{(k)}(t) + i2\pi\phi^{(k)}(t)}{i2\pi(k-1)!} \frac{V_f^{t^{k-1}g}(t,\eta)}{V_f^g(t,\eta)} \quad (5)$$

In this expansion, $V_f^g(t, \eta)$ denotes the STFT representation of signal $f(t)$. At this stage, the noisy signal $X(t, f)$ can be represented as the time-frequency domain signal:

$$X(t, f) = S(t, f) + N(t, f) \quad (6)$$

Where $S(t, f)$ denotes the signal and $N(t, f)$ is the noise added to the original signal. The main purpose of this research work is to estimate the signal $\hat{s}(t)$ from the corrupted noisy signal by aiming at decreasing the error between the original and reconstructed signal. The squared error can be expressed as:

$$E = \|\hat{s}(t) - s(t)\|_2^2 \quad (7)$$

Here, $s(t)$ and $\hat{s}(t)$ denote the actual and estimated signal in the time domain, which can be obtained by applying the inverse transform of STFT. This is obtained by the following equation:

$$\hat{s}(t, f) = M(t, f)X(t, f) \quad (8)$$

In Eq. (8), $M(t, f)$ represents a non-linear function which is used for mapping the $X(t, f)$ to a time-frequency representation of the estimated signal.

This problem is regarded as a machine learning problem that can be solved using a supervised learning approach. The current advancement of deep learning has motivated the adoption of the deep learning approach for the supervised learning algorithm. This work has established a deep learning-based model to learn the time-frequency domain features and construct a non-linear mapping using a training set.

### 3.3. DAE based Attention aware Skip Connection ECG filtering (DAE Based-ASCNet for ECG filtering)

This section presents the proposed solution for ECG denoising using the CNN-based Deep learning model. This model uses DAE modules, processing data by applying encoder and decoder modules. Moreover, an improved ReLU and a novel skip connection module are also presented, which mainly focus on increasing the outcome of the denoising model against strong noises.

The novelty of the proposed approach can be summarised with these comments: The traditional methods of Deep learning-based model use ReLU activation functions which suffer from information loss in the network layers. Similarly, the key features vanish when the conventional skip connection module is applied while mapping the features from the encoder to the decoder. Further, an attention mechanism is also incorporated to improve feature learning.

The proposed architecture uses improved ReLU, modified skip connection embedding and attention mechanism to improve the filtering performance. Below given figure 3 demonstrates the proposed DAE-ASCNet architecture.

The complete architecture is arranged into two models an encoder and a decoder module. In this encoding process, the 16x1 is set as kernel size for the initial two layers because the study presented in [] has proven the significance of large kernel size to the initial few convolutional layers to remove the baseline drift. Similarly, the reconstruction kernel size is also considered as 16. However, directly performing reconstruction from skip connection, deconvolution, ReLU, and Batch normalisation suffer from overfitting and accuracy-related issues; thus, an attention module is also assimilated to enhance the learning performance. The proposed attention module helps to extract the intra and inter-channel attention maps.

Moreover, the complete architecture uses an improved ReLU activation function, improving the non-linear transformation flexibility. This model also uses a modified skip connection model, which is used to connect the encoder and decoder layers. This helps retain the features and combines the low-level features with high-level features, minimising the loss.

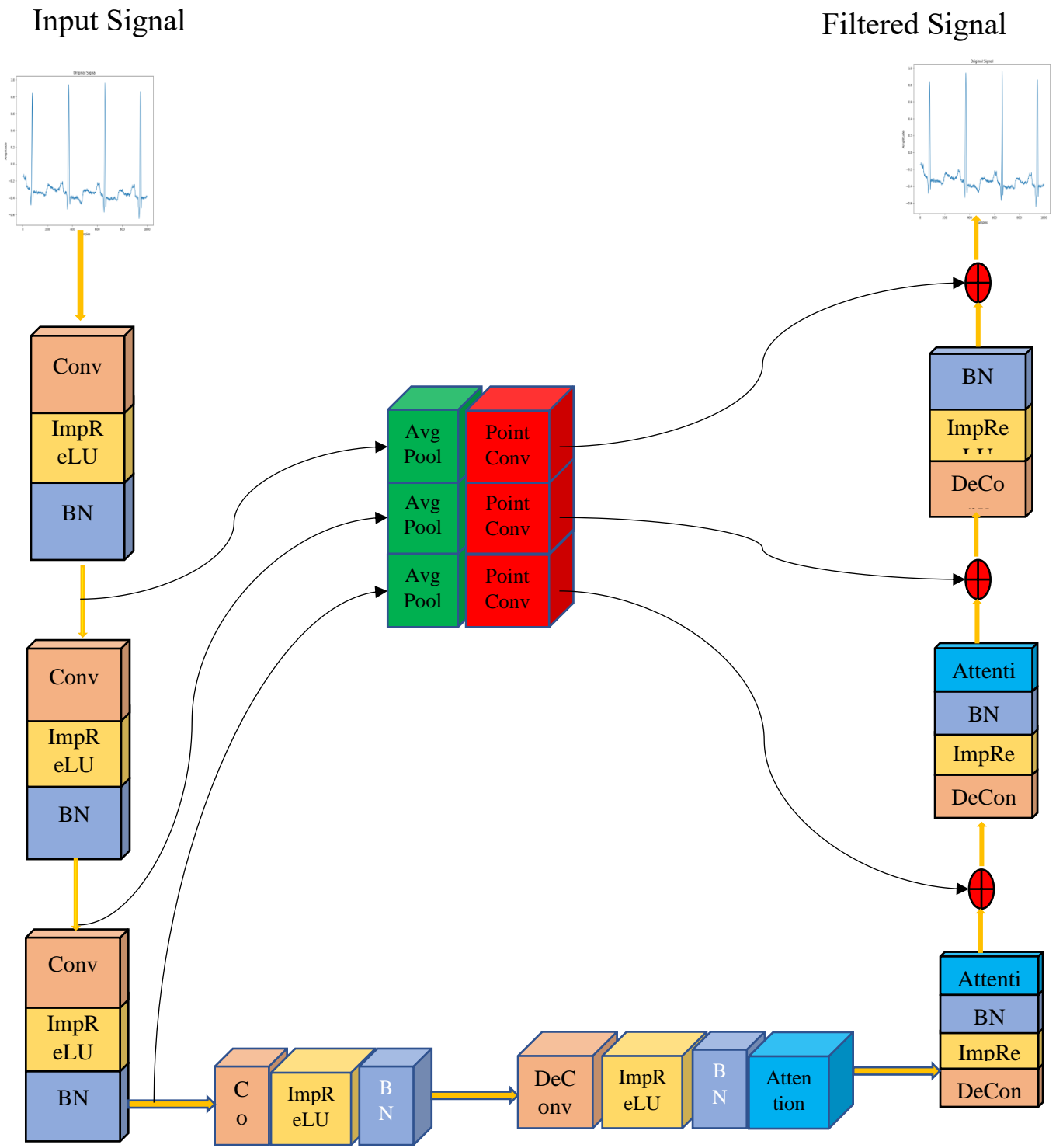

**Fig. 3 Proposed DAE-ASCNet for ECG filtering**

### 3.3.1. Improved ReLU

Generally, traditional architectures use linear rectifier units (ReLU) as an activation function. However, these activation functions are identical to each layer, limiting its ability to learn the attributes. Due to this, handling the noise with its dynamic nature becomes a challenging task. To overcome these issues, the modified activation function is presented. According to this activation function, the input signal is partitioned into positive and negative parts. After partition, the average pooled value of both parts is calculated to obtain the 1D vectors. These vectors are combined together and processed through the fully connected network. The complete process of improved ReLU is depicted below given figure 4.

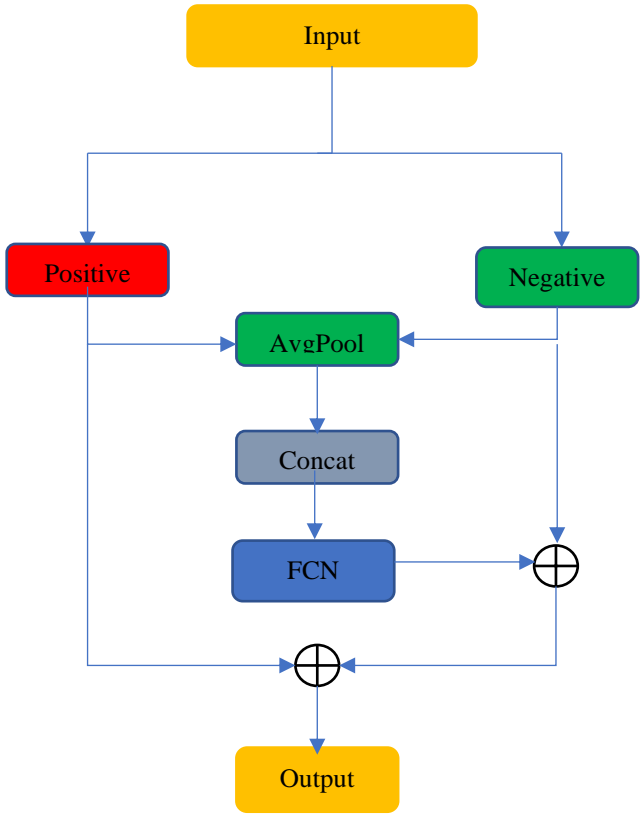

**Fig. 4 Architecture of modified ReLU**

The fully connected layer produces the slope of the negative part as $\alpha$. Finally, the output of this module can be expressed as:

$$y = \max(x, 0) + \alpha \times \min(x, 0) \quad (9)$$

Here, $x$ and $y$ represent the input and output of the module. This process helps to regulate the functioning of ReLU according to input and generates the corresponding features.

*3.3.2. Skip Connection*

This segment describes the skip connection model. The skip connection plays an important role in deep learning-based denoising architectures. As discussed before, the encoder layer generates feature maps and maps them to low dimensions. However, this process vanishes key features due to deepening the network. Thus, to overcome this issue, a novel skip connection module is presented, which considers encoder and decoder modules which perform average pooling and convolution operations on the input obtained from the encoder layer. Below given figure 5 depicts the skip connection architecture between two encoder and decoder layers.

The skip connection is expressed as:

$$\hat{a}_{0,n} = \frac{1}{C} \sum_{i=0}^{C-1} a_{i,n}$$

$$\hat{a} = [\hat{a}_{0,0}, \dots, \hat{a}_{0,N-1}]$$

$$a = \begin{bmatrix} a_{0,0} & \dots & a_{o,N-1} \\ \vdots & \ddots & \vdots \\ a_{C-1,0} & \dots & a_{C-1,N-1} \end{bmatrix} \quad (10)$$

Here $a$ denotes the input and $\hat{a}$ denotes the output from the skip connection module. The average pooling operation, which is applied on the output of the encoder, averages channel into one channel ($\hat{a}$). Further, convolution is applied to this data and fed to the decoder layer.

*3.3.3. Attention Module*

The attention mechanism plays an important role in deep learning-based schemes. The purpose of the attention mechanism is to enable the decoder to use the essential portions of the input. This is achieved by combining all of the encoded input vectors in a weighted fashion, with the essential vectors receiving the highest weights, i.e. attention helps to use the portion of the signal with the highest weights.

The proposed attention module contains channel and spatial attention modules to exploit the inter and intra-channel attributes. The figure 6 depicts the architecture of the attention module where channel and attention modules are presented.

In this stage, the output of the BN layer is processed through the max pooling and average pooling layers. The outcome of these layers is fed to the FCN layer. In the channel attention module, the outcome of FCN is added and processed through the sigmoid function to generate channel attention.

On the other hand, channel max pooling and average channel pooling are used, and the outcome of these two layers is concatenated to generate a single vector. This concatenated vector is processed through the Conv and Sigmoid layers to generate spatial attention, which generates the 2D feature map. The outcome of the attention mechanism can be expressed as:

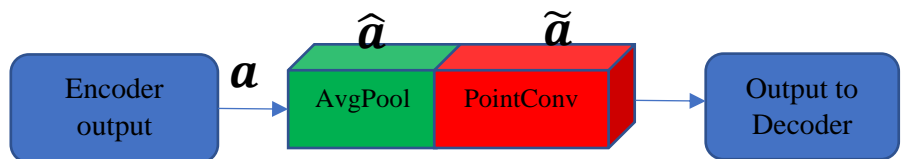

**Fig. 5 Skip connection module**

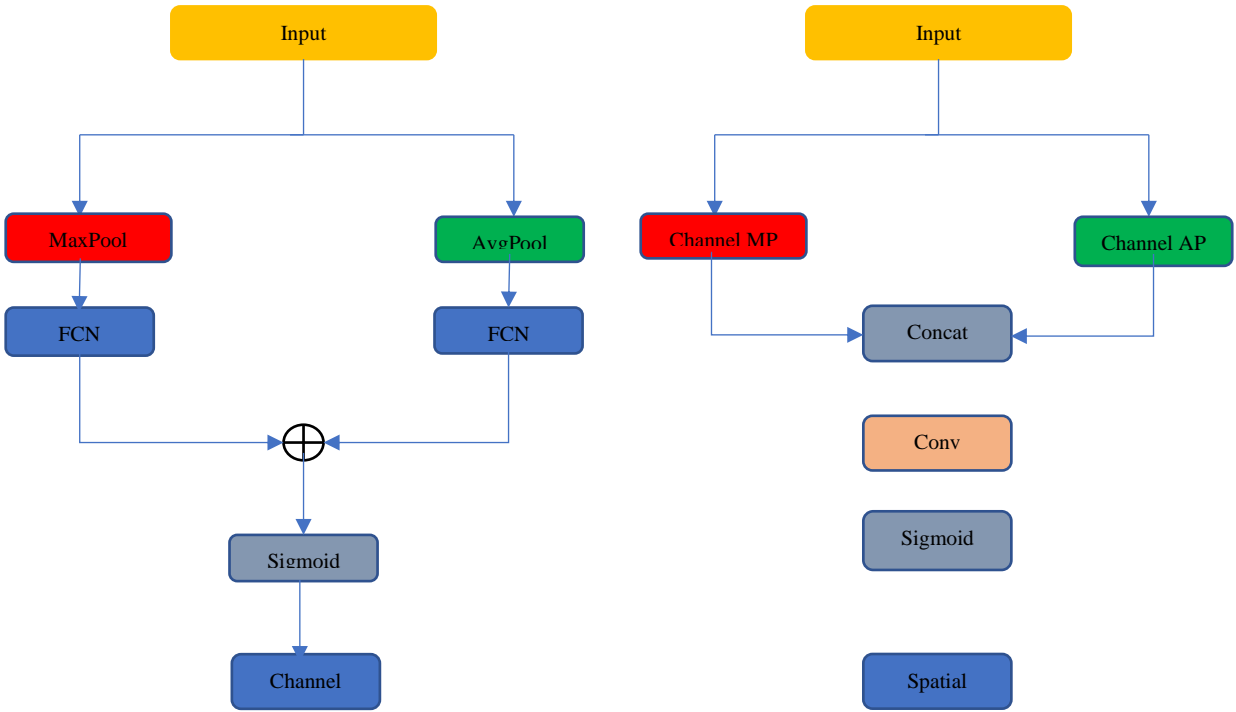

**Fig. 6 Channel and spatial attention modules**

$$\mu = \alpha_c \otimes x$$
$$y = \alpha_s \otimes \mu \quad (11)$$

Here, $x$ and $y$ denotes the input and output of the attention module, $\mu$ represents the intermediate variance, $\alpha_c$ and $\alpha_s$ denotes the channel and spatial attention module.

## 4. Results and Discussion

This segment describes the outcome of the proposed approach. It presents a comparative analysis where the performance of the proposed ASCNet approach is compared with prevailing techniques to show the robustness of the proposed approach. The proposed approach is implemented on the publicly available MIT-BIH dataset. The performance of the ASCNet methodology is realised in terms of improved SNR, Mean Squared Error, and percent root mean square difference (PRD) below given sections 4.1 and 4.2. describes the dataset details and performance measurement parameters, respectively.

### 4.1. Dataset Details

The MIT-BIH dataset encompasses 48 hours of recording two-channel ambulatory ECG signals obtained from 47 subjects. Out of these 47 subjects. Out of these 48 hours, 23 readings are chosen from 24 hours' ambulatory signals, including the mixed population of inpatient and outpatient as 60% and 40% subjects, respectively. The remaining 25 readings are obtained from the same signal set, including clinically vital arrhythmias. For simplification, these signals were processed through the digitisation phase, which generates the digitised signal at 360 samples per second per channel. This digitisation is done with an 11-bit resolution over a 10mV range. Expert cardiologists annotate these signals, and the entire dataset contains 110,000 annotations. These datasets are freely available [24]. Some of the samples of these signals are depicted in figure 7.

### 4.2. Performance Measurement

This segment describes the performance measurement parameters which are used in this article to realise the performance of the proposed model.

- SNR: it is used to measure the quality of the reconstructed signal. This can be computed as:

$$SNR_{imp} = 10 \log_{10}\left(\frac{\sum_{n=1}^{N}(f_i)^2}{\sum_{n=1}^{N}|f(n) - u(n)|^2}\right)$$

- Means Squared Error (MSE): this matrix denotes the average squared difference between the actual ECG signal and filtered ECG signal. This can be expressed as:

$$MSE = \frac{1}{N}\sum_{n=1}^{N}[f(n) - u(n)]^2$$

Based on this MSE, the Root Mean Square value is computed as $RMSE = \sqrt{\frac{1}{N}\sum_{n=1}^{N}[f(n) - u(n)]^2}$

### 4.3. Comparative Analysis

This section presents a detailed discussion on the measurement of filtering performance. This work has considered several types of noises, such as white noise, baseline wander noise, motion artifacts, and electrode motion noise. These noises are added to the original signal at different SNR levels. Below given figure 8 depicts some samples of the original signal, different noise types and combined signals with different noises.

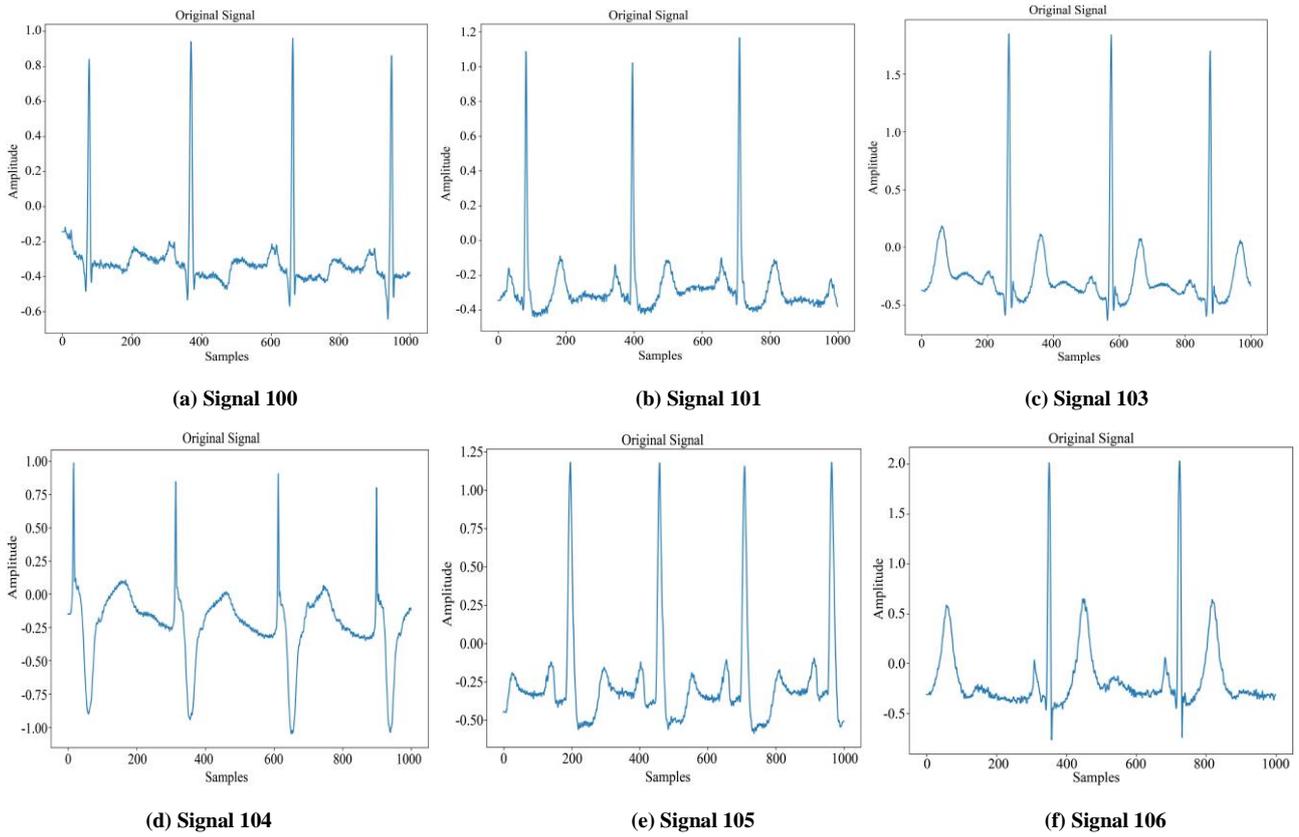

**Fig. 7 Sample ECG signals obtained from MIT-BIH dataset**

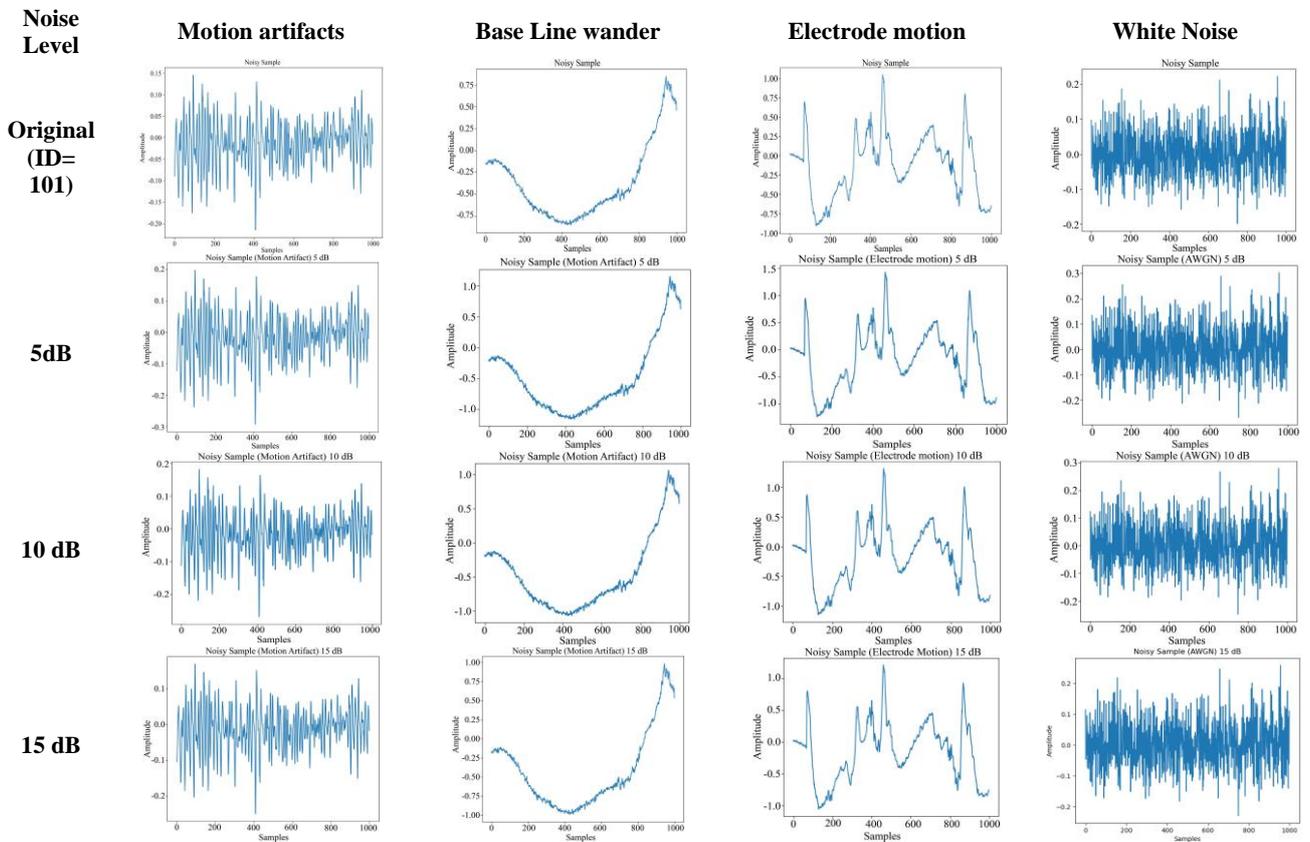

**Fig. 8 Sample ECG signals obtained from MIT-BIH dataset**

Table 2. SNR and MSE performance comparison for AWGN

|  | Noise Level | Parameter | Record Number | | |
|---|---|---|---|---|---|
|  |  |  | 100m | 103m | 213m |
| BP-ADMM | 5 dB | SNR | 14.71 | 14.86 | 13.92 |
|  |  | MSE | 0.0031 | 0.0041 | 0.024 |
|  | 15 dB | SNR | 15.15 | 16.10 | 14.35 |
|  |  | MSE | 0.0021 | 0.0021 | 0.0031 |
| Wavelet | 5 dB | SNR | 5.312 | 4.85 | 7.49 |
|  |  | MSE | 0.0247 | 0.085 | 0.125 |
|  | 15 dB | SNR | 9.7863 | 4.85 | 7.49 |
|  |  | MSE | 0.1002 | 0.086 | 0.127 |
| Average | 5 dB | SNR | 2.3397 | 2.3837 | 3.0389 |
|  |  | MSE | 0.5095 | 0.5180 | 1.1890 |
|  | 15 dB | SNR | 3.2940 | 3.3015 | 3.4546 |
|  |  | MSE | 0.2670 | 0.3080 | 0.9620 |
| TV | 5 dB | SNR | 4.6338 | 4.8580 | 9.2792 |
|  |  | MSE | 0.3030 | 0.3070 | 0.2840 |
|  | 15 dB | SNR | 10.725 | 11.28 | 12.750 |
|  |  | MSE | 0.0480 | 0.0490 | 0.0715 |
| Proposed | 5 dB | SNR | 21.220 | 18.96 | 22.30 |
|  |  | MSE | 0.0011 | 0.0018 | 0.0015 |
|  | 15 dB | SNR | 24.30 | 26.22 | 25.30 |
|  |  | MSE | 0.0016 | 0.0014 | 0.0011 |

Further, the performance of the proposed approach is compared with wavelet, Average filtering, Total variation and BP-ADMM filtering techniques in terms of SNR.

Table 3. SNR performance for 5dB white noise

| ECG sample | Wavelet | Average | TV | BP-ADMM | Proposed |
|---|---|---|---|---|---|
| 105 | 5.474 | 2.4290 | 4.9170 | 6.3298 | 15.2025 |
| 106 | 5.5671 | 2.433 | 5.4115 | 5.8936 | 18.1076 |
| 107 | 11.3446 | 3.2118 | 11.6603 | 11.67 | 19.3310 |
| 108 | 4.4446 | 2.2665 | 4.2473 | 4.8901 | 22.1501 |
| 109 | 7.798 | 2.8677 | 7.6668 | 8.1662 | 24.2210 |
| 111 | 3.0188 | 1.8348 | 2.7679 | 3.6756 | 18.5041 |
| 112 | 11.7158 | 3.2256 | 11.432 | 12.0141 | 17.5061 |
| 113 | 6.647 | 2.7343 | 6.4469 | 6.7013 | 19.5511 |
| 114 | 2.9079 | 1.739 | 2.2118 | 3.5163 | 20.2140 |
| 115 | 7.785 | 2.8554 | 7.5518 | 7.9067 | 18.5013 |
| 116 | 13.23 | 3.3241 | 13.0733 | 13.6521 | 19.205 |
| 117 | 11.26 | 3.1911 | 10.62 | 11.8165 | 21.3015 |
| 118 | 12.15 | 3.2713 | 11.8001 | 13.1671 | 22.3058 |
| 119 | 12.4043 | 3.2694 | 11.8452 | 12.9172 | 24.3014 |
| 200 | 6.3196 | 2.6159 | 6.0043 | 6.7027 | 21.0257 |
| 201 | 3.6453 | 1.9371 | 3.2885 | 4.0273 | 22.0158 |
| 202 | 2.9598 | 1.7462 | 2.4286 | 3.7509 | 23.210 |
| 203 | 6.0959 | 2.6427 | 5.8747 | 6.9253 | 22.105 |
| 205 | 5.8739 | 2.4731 | 5.322 | 6.0851 | 23.015 |
| 205 | 5.0551 | 2.405 | 4.8458 | 5.6361 | 21.05 |
| 208 | 7.4163 | 2.8221 | 7.4943 | 8.1552 | 20.134 |
| 209 | 3.6206 | 1.9473 | 3.2982 | 4.7622 | 18.5621 |
| 210 | 3.5107 | 2.0053 | 3.2245 | 4.1386 | 17.2514 |
| 212 | 4.753 | 2.2869 | 4.5983 | 5.126 | 24.3041 |
| 231 | 3.9913 | 2.0872 | 3.6602 | 4.5625 | 26.3018 |
| 232 | 3.0207 | 1.7841 | 2.2933 | 3.6549 | 22.5017 |
| 233 | 8.7804 | 2.9848 | 8.5975 | 9.2172 | 26.3017 |
| 234 | 4.0129 | 2.1281 | 3.6201 | 4.5516 | 25.1134 |
| Mean SNR | 6.6006 | 2.5185 | 6.3037 | 7.129 | 21.189 |

Similarly, this dataset contains some standard noisy signals that generally contaminate the quality of the original signal during acquisition. Below given figure depicts the samples of the original noisy sample along with noisy samples with different levels of SNR.

These noise samples are added to the original signal to incorporate real-time noise scenarios. This task of adding noise helps to evaluate the robustness of the proposed approach.

*4.3.1. Filtering performance Comparison for White noise*

Based on these experiments, the performance of the proposed approach is compared with other existing schemes, as mentioned in [25]. In [25], several other ECG filtering techniques, such as BP-ADMM, Wavelet, Average filtering, and total variation (TV), are evaluated. Below given table shows a comparative analysis of this experiment for varied levels of White noise.

*4.3.2. Filtering performance comparison for MA, EM and BW noise*

Similarly, Xu et al. [26] presented a new approach combining Generative Adversarial Network and Residual Network. In [26], authors have evaluated the performance of state-of-art ECG filtering algorithms such as S-transform, wavelet transforms, and stacked and improved denoising autoencoder algorithms. The performance of the proposed approach is compared with these techniques, where different noise types such as MA, EM, and BW are added at 0dB, 1.25 dB and 5dB noise levels. Below given table 3 shows the comparative analysis where the performance of these techniques is measured in terms of RMSE and SNR for MA noise.

Table 4. SNR and RMSE performance for MA noise

| | | | 103m | 105m | 111m | 116m | 122 | 205 | 213 | 219 | 223 | 230 |
|---|---|---|---|---|---|---|---|---|---|---|---|---|
| WT | 0 dB | SNR | 19.70 | 22.10 | 20.00 | 12.40 | 6.70 | 21.20 | 11.80 | 7.30 | 18.50 | 18.00 |
| | | RMSE | 0.045 | 0.04 | 0.051 | 0.1100 | 0.180 | 0.045 | 0.095 | 0.150 | 0.050 | 0.070 |
| | 1.25 dB | SNR | 16.90 | 22.50 | 19.70 | 14.45 | 7.45 | 20.20 | 13.20 | 8.65 | 20.12 | 18.80 |
| | | RMSE | 0.065 | 0.040 | 0.050 | 0.085 | 0.160 | 0.045 | 0.080 | 0.120 | 0.041 | 0.065 |
| | 5 dB | SNR | 15.80 | 24.10 | 18.80 | 19.10 | 11.15 | 16.50 | 18.90 | 14.60 | 21.40 | 21.10 |
| | | RMSE | 0.067 | 0.032 | 0.057 | 0.046 | 0.109 | 0.075 | 0.042 | 0.065 | 0.041 | 0.046 |
| ST | 0 dB | SNR | 10.38 | 10.11 | 8.19 | 8.22 | 9.2 | 8.29 | 8.82 | 10.00 | 9.89 | 8.7 |
| | | RMSE | 0.302 | 0.315 | 0.391 | 0.391 | 0.350 | 0.380 | 0.360 | 0.315 | 0.320 | 0.365 |
| | 1.25 dB | SNR | 10.90 | 10.40 | 8.60 | 8.50 | 9.70 | 8.60 | 9.70 | 10.60 | 10.60 | 9.10 |
| | | RMSE | 0.290 | 0.300 | 0.370 | 0.380 | 0.330 | 0.370 | 0.330 | 0.290 | 0.300 | 0.350 |
| | 5 dB | SNR | 12.60 | 12.80 | 9.90 | 9.80 | 11.70 | 9.90 | 12.50 | 12.90 | 13.40 | 10.20 |
| | | RMSE | 0.230 | 0.231 | 0.310 | 0.320 | 0.261 | 0.321 | 0.235 | 0.220 | 0.210 | 0.310 |
| S-DAE | 0 dB | SNR | 18.90 | 22.90 | 22.9 | 17.90 | 17.80 | 20.10 | 18.20 | 16.20 | 20.30 | 21.10 |
| | | RMSE | 0.045 | 0.035 | 0.035 | 0.055 | 0.050 | 0.051 | 0.045 | 0.050 | 0.050 | 0.050 |
| | 1.25 dB | SNR | 19.10 | 23.35 | 22.90 | 18.50 | 18.10 | 20.13 | 18.75 | 17.70 | 21.20 | 21.20 |
| | | RMSE | 0.048 | 0.038 | 0.039 | 0.049 | 0.045 | 0.050 | 0.045 | 0.045 | 0.041 | 0.046 |
| | 5 dB | SNR | 19.32 | 24.15 | 22.96 | 20.65 | 21.74 | 20.15 | 20.18 | 20.15 | 23.722 | 21.353 |
| | | RMSE | 0.042 | 0.035 | 0.036 | 0.039 | 0.0329 | 0.052 | 0.034 | 0.034 | 0.030 | 0.045 |
| I-DAE | 0 dB | SNR | 21.40 | 24.70 | 23.20 | 19.20 | 19.60 | 24.20 | 19.60 | 18.79 | 22.89 | 22.61 |
| | | RMSE | 0.031 | 0.031 | 0.034 | 0.041 | 0.041 | 0.029 | 0.037 | 0.038 | 0.031 | 0.0358 |
| | 1.25 dB | SNR | 22.45 | 24.87 | 23.37 | 20.82 | 20.12 | 24.50 | 19.79 | 19.60 | 23.40 | 22.62 |
| | | RMSE | 0.032 | 0.031 | 0.033 | 0.039 | 0.040 | 0.033 | 0.036 | 0.033 | 0.031 | 0.0369 |
| | 5 dB | SNR | 23.32 | 25.12 | 23.30 | 22.40 | 20.60 | 24.65 | 20.63 | 21.97 | 24.21 | 22.60 |
| | | RMSE | 0.025 | 0.025 | 0.033 | 0.030 | 0.035 | 0.035 | 0.030 | 0.033 | 0.025 | 0.035 |
| PM [26] | 0 dB | SNR | 37.20 | 34.05 | 31.81 | 34.90 | 43.20 | 42.87 | 34.21 | 34.02 | 35.65 | 36.40 |
| | | RMSE | 0.0051 | 0.0031 | 0.0135 | 0.0052 | 0.0009 | 0.0076 | 0.0104 | 0.0090 | 0.0085 | 0.0227 |
| | 1.25 dB | SNR | 32.30 | 29.06 | 26.94 | 29.25 | 39.62 | 31.50 | 33.29 | 29.23 | 29.06 | 31.36 |
| | | RMSE | 0.0137 | 0.0165 | 0.0165 | 0.0141 | 0.0165 | 0.0035 | 0.0205 | 0.0115 | 0.0140 | 0.0295 |
| | 5 dB | SNR | 59.57 | 56.45 | 53.65 | 56.71 | 51.60 | 58.60 | 58.90 | 57.72 | 63.85 | 59.60 |
| | | RMSE | 0.0155 | 0.0004 | 0.0070 | 0.0005 | 0.0061 | 0.0010 | 0.0091 | 0.0055 | 0.0038 | 0.015 |
| PS | 0 dB | SNR | 41.20 | 38.56 | 36.50 | 44.20 | 46.20 | 52.20 | 39.50 | 39.50 | 39.45 | 45.62 |
| | | RMSE | 0.0039 | 0.0025 | 0.011 | 0.0035 | 0.0008 | 0.0066 | 0.010 | 0.0088 | 0.0075 | 0.0212 |
| | 1.25 dB | SNR | 38.92 | 35.20 | 35.62 | 36.51 | 43.56 | 39.21 | 39.50 | 36.21 | 35.62 | 39.65 |
| | | RMSE | 0.0112 | 0.0151 | 0.0151 | 0.0132 | 0.0115 | 0.0031 | 0.0185 | 0.0110 | 0.0130 | 0.0154 |
| | 5 dB | SNR | 63.20 | 66.21 | 63.25 | 62.20 | 56.25 | 66.20 | 66.20 | 66.32 | 66.90 | 65.201 |
| | | RMSE | 0.0110 | 0.0003 | 0.0061 | 0.0004 | 0.0055 | 0.0008 | 0.0085 | 0.0049 | 0.0028 | 0.0109 |

In the next experiment, the EM noise is considered added to the 103m, 105m, 111m, 116m, 122m,205m,213m,219m,223m, and 230m signals of the MIT-BIH dataset. This noise is added at 0 dB, 1.25 dB, and 5 dB levels. Below given table shows the comparative performance of EM noise filtering.

**Table 5. SNR and RMSE performance for EM noise**

|   |   |   | 103m | 105m | 111m | 116m | 122 m | 205 m | 213 m | 219 m | 223 m | 230 m |
|---|---|---|---|---|---|---|---|---|---|---|---|---|
| WT | 0 dB | SNR | 9.50 | 21.47 | 9.35 | 9.60 | 9.70 | 18.30 | 15.10 | 13.20 | 18.10 | 11.90 |
|  |  | RMSE | 0.135 | 0.043 | 0.170 | 0.135 | 0.130 | 0.065 | 0.065 | 0.075 | 0.055 | 0.130 |
|  | 1.25 dB | SNR | 10.35 | 23.10 | 9.66 | 10.55 | 9.50 | 20.31 | 15.85 | 13.80 | 21.08 | 12.93 |
|  |  | RMSE | 0.130 | 0.040 | 0.170 | 0.120 | 0.130 | 0.051 | 0.060 | 0.071 | 0.041 | 0.118 |
|  | 5 dB | SNR | 13.08 | 28.40 | 14.95 | 13.75 | 8.70 | 21.36 | 19.21 | 16.25 | 21.10 | 15.89 |
|  |  | RMSE | 0.090 | 0.022 | 0.092 | 0.085 | 0.145 | 0.042 | 0.039 | 0.055 | 0.030 | 0.085 |
| ST | 0 dB | SNR | 6.39 | 6.15 | 5.50 | 5.50 | 5.90 | 5.60 | 5.83 | 6.15 | 6.18 | 6.30 |
|  |  | RMSE | 0.475 | 0.490 | 0.530 | 0.530 | 0.510 | 0.530 | 0.511 | 0.490 | 0.490 | 0.485 |
|  | 1.25 dB | SNR | 7.58 | 7.40 | 6.41 | 6.35 | 6.95 | 6.50 | 7.16 | 7.18 | 7.46 | 7.49 |
|  |  | RMSE | 0.421 | 0.428 | 0.475 | 0.482 | 0.445 | 0.471 | 0.441 | 0.435 | 0.421 | 0.421 |
|  | 5 dB | SNR | 10.31 | 10.35 | 8.55 | 8.30 | 9.65 | 8.50 | 10.12 | 10.04 | 10.71 | 10.40 |
|  |  | RMSE | 0.304 | 0.303 | 0.379 | 0.380 | 0.330 | 0.370 | 0.315 | 0.315 | 0.289 | 0.299 |
| S-DAE | 0 dB | SNR | 18.94 | 23.45 | 22.33 | 19.18 | 17.87 | 20.08 | 19.20 | 17.53 | 22.65 | 20.79 |
|  |  | RMSE | 0.046 | 0.035 | 0.035 | 0.045 | 0.046 | 0.049 | 0.040 | 0.044 | 0.031 | 0.045 |
|  | 1.25 dB | SNR | 19.07 | 23.82 | 22.43 | 19.69 | 18.95 | 20.11 | 19.75 | 18.30 | 23.20 | 20.91 |
|  |  | RMSE | 0.047 | 0.035 | 0.040 | 0.045 | 0.045 | 0.05 | 0.04 | 0.04 | 0.030 | 0.045 |
|  | 5 dB | SNR | 19.25 | 24.55 | 22.66 | 21 | 21.15 | 20.24 | 20.98 | 20.08 | 21.39 | 21.40 |
|  |  | RMSE | 0.040 | 0.030 | 0.036 | 0.036 | 0.032 | 0.048 | 0.031 | 0.033 | 0.027 | 0.044 |
| I-DAE | 0 dB | SNR | 22.70 | 23.70 | 23.40 | 21.34 | 17.70 | 23.47 | 19.30 | 18.40 | 23.20 | 22.45 |
|  |  | RMSE | 0.028 | 0.033 | 0.034 | 0.05 | 0.050 | 0.033 | 0.040 | 0.041 | 0.031 | 0.038 |
|  | 1.25 dB | SNR | 22.95 | 23.94 | 23.57 | 21.82 | 18.75 | 23.55 | 19.80 | 19.10 | 23.55 | 22.54 |
|  |  | RMSE | 0.030 | 0.031 | 0.032 | 0.032 | 0.041 | 0.031 | 0.034 | 0.035 | 0.037 | 0.035 |
|  | 5 dB | SNR | 23.40 | 24.66 | 23.65 | 23.08 | 20.81 | 23.66 | 20.69 | 21.01 | 24 | 22.81 |
|  |  | RMSE | 0.025 | 0.030 | 0.027 | 0.030 | 0.033 | 0.030 | 0.035 | 0.030 | 0.028 | 0.037 |
| PM [26] | 0 dB | SNR | 28.67 | 27.56 | 24.25 | 28.57 | 35.57 | 35.96 | 30.81 | 26.73 | 25.036 | 25.045 |
|  |  | RMSE | 0.012 | 0.0344 | 0.024 | 0.0117 | 0.0135 | 0.0079 | 0.0314 | 0.0110 | 0.0125 | 0.0215 |
|  | 1.25 dB | SNR | 35.46 | 31.24 | 27.08 | 32.89 | 39.53 | 35.52 | 30.23 | 29.882 | 31.05 | 35.77 |
|  |  | RMSE | 0.0085 | 0.0035 | 0.0166 | 0.0110 | 0.0135 | 0.0044 | 0.0193 | 0.0114 | 0.0139 | 0.0186 |
|  | 5 dB | SNR | 62.71 | 53.75 | 57.02 | 59.02 | 64.46 | 67.39 | 58.73 | 60.07 | 55.32 | 60.82 |
|  |  | RMSE | 0.0030 | 0.0102 | 0.022 | 0.0229 | 0.0077 | 0.0107 | 0.0011 | 0.0100 | 0.0010 | 0.0178 |
| PS | 0 dB | SNR | 35.12 | 32.25 | 32.10 | 32.20 | 41.20 | 42.13 | 36.20 | 32.25 | 35.61 | 35.112 |
|  |  | RMSE | 0.009 | 0.031 | 0.018 | 0.011 | 0.0112 | 0.0035 | 0.0251 | 0.01 | 0.0112 | 0.0112 |
|  | 1.25 dB | SNR | 41.23 | 39.52 | 32.25 | 39.15 | 46.20 | 42.30 | 35.20 | 36.10 | 36.50 | 41.205 |
|  |  | RMSE | 0.0044 | 0.0022 | 0.0106 | 0.011 | 0.0151 | 0.0028 | 0.0115 | 0.010 | 0.0121 | 0.0118 |
|  | 5 dB | SNR | 68.91 | 62.30 | 65.32 | 62.30 | 73.25 | 76.50 | 63.21 | 63.25 | 62.30 | 75.20 |
|  |  | RMSE | 0.0028 | 0.010 | 0.018 | 0.0215 | 0.0062 | 0.01 | 0.001 | 0.0100 | 0.001 | 0.0112 |

Finally, in this investigation, BW noise is considered, which is added to the original signal at varied levels, such as 0dB, 1.25dB, and 5dB. Below given table shows the comparative performance for BW noise filtering.

**Table 6. SNR and RMSE performance for BW noise**

|  |  |  | 103m | 105m | 111m | 116m | 122 | 205 | 213 | 219 | 223 | 230 |
|---|---|---|---|---|---|---|---|---|---|---|---|---|
| WT | 0 dB | SNR | 14.875 | 31.50 | 18.40 | 20.10 | 9.15 | 22.60 | 20.85 | 18.70 | 17.30 | 22.20 |
|  |  | RMSE | 0.075 | 0.015 | 0.062 | 0.043 | 0.138 | 0.037 | 0.35 | 0.041 | 0.062 | 0.042 |
|  | 1.25 dB | SNR | 14.85 | 31.90 | 18.40 | 20.05 | 8.56 | 22.73 | 20.50 | 20.21 | 17.35 | 22.20 |
|  |  | RMSE | 0.0734 | 0.0129 | 0.0599 | 0.0419 | 0.145 | 0.032 | 0.035 | 0.35 | 0.060 | 0.040 |
|  | 5 dB | SNR | 14.85 | 32.60 | 18.40 | 20.50 | 8.20 | 22.90 | 19.10 | 21.45 | 17.45 | 22.15 |
|  |  | RMSE | 0.075 | 0.011 | 0.065 | 0.041 | 0.150 | 0.035 | 0.041 | 0.028 | 0.062 | 0.040 |
| ST | 0 dB | SNR | 11.4 | 11.50 | 9.20 | 9.10 | 1060 | 9.30 | 11.55 | 11.50 | 12.14 | 11.68 |
|  |  | RMSE | 0.269 | 0.265 | 0.346 | 0.355 | 0.296 | 0.342 | 0.265 | 0.264 | 0.45 | 0.260 |
|  | 1.25 dB | SNR | 12.05 | 12.20 | 9.60 | 9.35 | 11.169 | 9.65 | 12.20 | 12.20 | 12.90 | 12.90 |
|  |  | RMSE | 0.245 | 0.245 | 0.330 | 0.340 | 0.275 | 0.325 | 0.245 | 0.245 | 0.2249 | 0.240 |
|  | 5 dB | SNR | 13.549 | 13.669 | 10.40 | 10.05 | 12.30 | 10.45 | 13.70 | 13.65 | 14.75 | 14.75 |
|  |  | RMSE | 0.22 | 0.206 | 0.301 | 0.316 | 0.241 | 0.3 | 0.206 | 0.208 | 0.108 | 0.202 |
| S-DAE | 0 dB | SNR | 20.35 | 24.90 | 23.05 | 18.80 | 19.45 | 20.05 | 19.42 | 19.25 | 22.90 | 20.50 |
|  |  | RMSE | 0.038 | 0.028 | 0.034 | 0.046 | 0.039 | 0.049 | 0.034 | 0.035 | 0.030 | 0.048 |
|  | 1.25 dB | SNR | 20.51 | 23.24 | 23.06 | 19.55 | 19.58 | 20.12 | 20.30 | 19.85 | 23.75 | 20.65 |
|  |  | RMSE | 0.038 | 0.030 | 0.030 | 0.044 | 0.037 | 0.049 | 0.035 | 0.034 | 0.028 | 0.045 |
|  | 5 dB | SNR | 20.76 | 25.45 | 23.02 | 21.28 | 21 | 20.33 | 21.33 | 21.115 | 25.40 | 21.03 |
|  |  | RMSE | 0.037 | 0.027 | 0.035 | 0.35 | 0.035 | 0.045 | 0.030 | 0.030 | 0.024 | 0.045 |
| I-DAE | 0 dB | SNR | 23.78 | 25.40 | 23.31 | 23.51 | 20.07 | 20.07 | 21.30 | 23.05 | 24.22 | 22.70 |
|  |  | RMSE | 0.025 | 0.025 | 0.035 | 0.026 | 0.048 | 0.049 | 0.049 | 0.031 | 0.025 | 0.037 |
|  | 1.25 dB | SNR | 23.80 | 25.40 | 23.30 | 23.60 | 20.10 | 20.10 | 21.35 | 23.30 | 24.40 | 22.75 |
|  |  | RMSE | 0.025 | 0.025 | 0.035 | 0.028 | 0.051 | 0.051 | 0.033 | 0.024 | 0.025 | 0.036 |
|  | 5 dB | SNR | 23.89 | 25.42 | 23.30 | 23.75 | 20.10 | 20.10 | 21.45 | 24.10 | 24.65 | 22.80 |
|  |  | RMSE | 0.024 | 0.026 | 0.034 | 0.025 | 0.048 | 0.048 | 0.032 | 0.022 | 0.025 | 0.004 |
| PM [26] | 0 dB | SNR | 29.75 | 29.66 | 28.35 | 29.79 | 38.25 | 28.50 | 33.08 | 27.30 | 30.17 | 33.81 |
|  |  | RMSE | 0.0041 | 0.0033 | 0.0091 | 0.0029 | 0.0031 | 0.0067 | 0.0088 | 0.0067 | 0.0067 | 0.0066 |
|  | 1.25 dB | SNR | 36.10 | 34.62 | 31.55 | 33.97 | 41.27 | 28.59 | 33.61 | 30.73 | 32.25 | 34.82 |
|  |  | RMSE | 0.0135 | 0.0032 | 0.0110 | 0.0087 | 0.1469 | 0.0183 | 0.0093 | 0.0075 | 0.0075 | 0.0069 |
|  | 5 dB | SNR | 67.79 | 62.17 | 63.47 | 62.20 | 69.24 | 70.65 | 70.07 | 64.33 | 68.27 | 71.37 |
|  |  | RMSE | 0.0013 | 0.0020 | 0.0007 | 0.0105 | 0.0004 | 0.0013 | 0.0006 | 0.0103 | 0.0007 | 0.0172 |
| PS | 0 dB | SNR | 31.22 | 33.20 | 31.20 | 33.51 | 41.20 | 32.20 | 36.20 | 36.51 | 39.51 | 39.55 |
|  |  | RMSE | 0.0035 | 0.0013 | 0.0089 | 0.0015 | 0.0022 | 0.0055 | 0.0012 | 0.0055 | 0.0012 | 0.0054 |
|  | 1.25 dB | SNR | 39.42 | 39.25 | 33.56 | 36.20 | 46.50 | 35.10 | 39.42 | 35.61 | 36.51 | 35.62 |
|  |  | RMSE | 0.0115 | 0.0011 | 0.0150 | 0.0051 | 0.161 | 0.0321 | 0.0090 | 0.0043 | 0.0060 | 0.0055 |
|  | 5 dB | SNR | 71.50 | 65.18 | 66.50 | 68.50 | 73.25 | 73.21 | 76.55 | 68.91 | 71.23 | 79.20 |
|  |  | RMSE | 0.0011 | 0.0018 | 0.0004 | 0.0112 | 0.0001 | 0.0010 | 0.0005 | 0.010 | 0.0005 | 0.0161 |

*4.3.3. Signal Reconstruction Performance Comparison for MA, EM and BW Noise*

Further, the SNR performance is measured for different noises, such as BW noise, EM noise, and MA noise for different ECG signals. The SNR measurement demonstrates the quality of signal reconstruction. In [26], Xu et al. considered 103, 105, 111, 116, 122, 205, 213, 219, 223, and 230 sample IDs to measure the performance. Below given figure depicts the filtering performance for EM noise with an SNR of 1.25 dB

For BW noise removal, the average SNR performance is obtained as 5.96 dB, 14.84dB, 18.11dB, 21.36dB, 31.43dB, and 34.91dB by using S-Transform, Wavelet Transform, Stacked DAE, Improved DAE, GAN [26], and Proposed approach, respectively.

Maximum SNR is obtained as 36.5dB by using the proposed approach.

In the next experiment, the proposed experiment is conducted for MA noise which is added to the original signal with 1.25dB. This performance is also measured for the same ECG samples which are mentioned in the previous experiment. Below given figure shows the comparative analysis for MA noise Reduction.

According to the above experiment, the average performance is obtained as 9.35dB, 15.32dB, 18.04dB, 21.03dB, 33.44dB, and 35.81dB by using S-Transform, Wavelet Transform, Stacked DAE, Improved DAE, GAN [26], and Proposed approach, respectively. Finally, the performance for EM noise reduction is measured, which is combined with the original signal with an SNR of 1.25 dB .

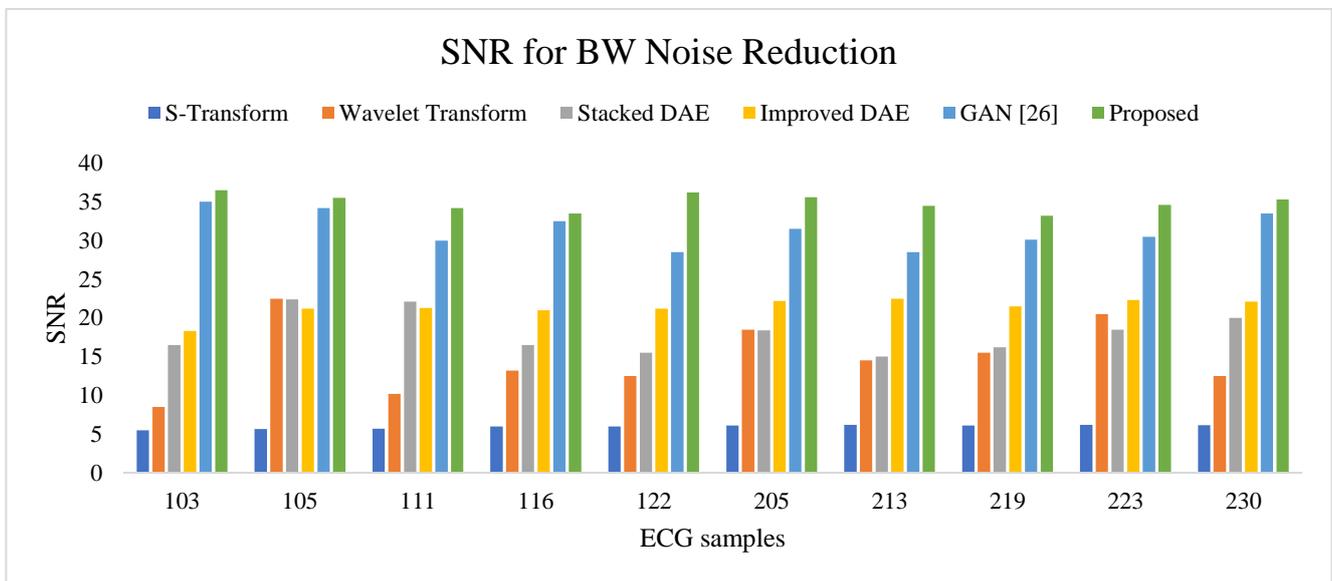

**Fig. 9 SNR performance for BW noise reduction with an SNR of 1.25 dB**

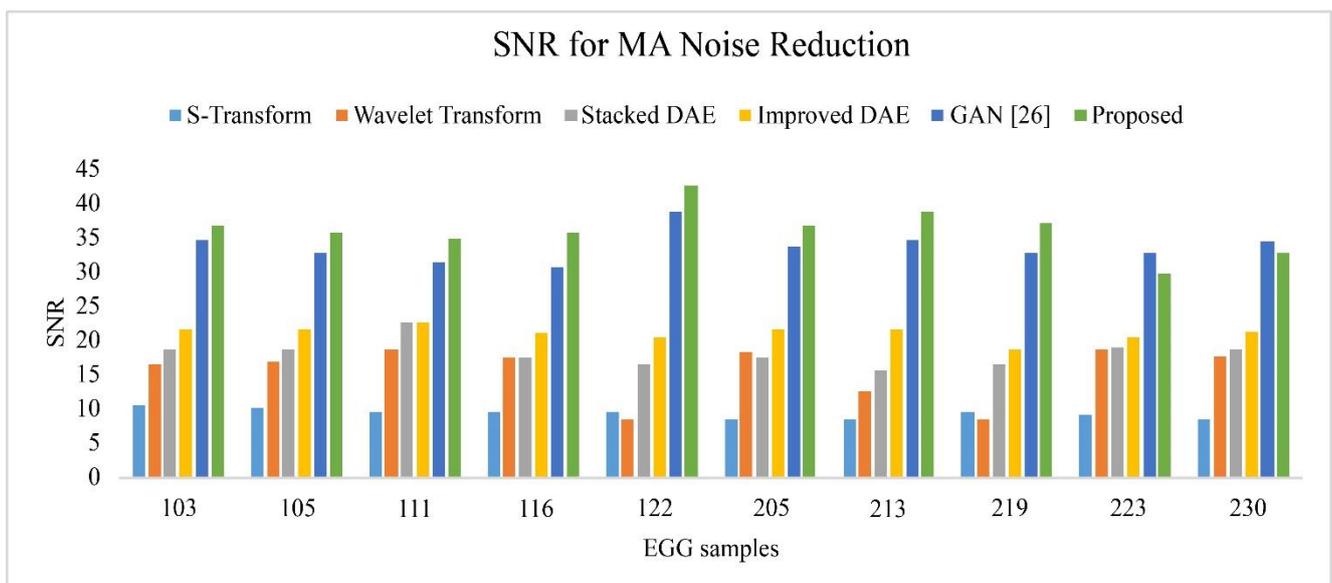

**Fig. 10 SNR performance for MA noise reduction with an SNR of 1.25 dB**

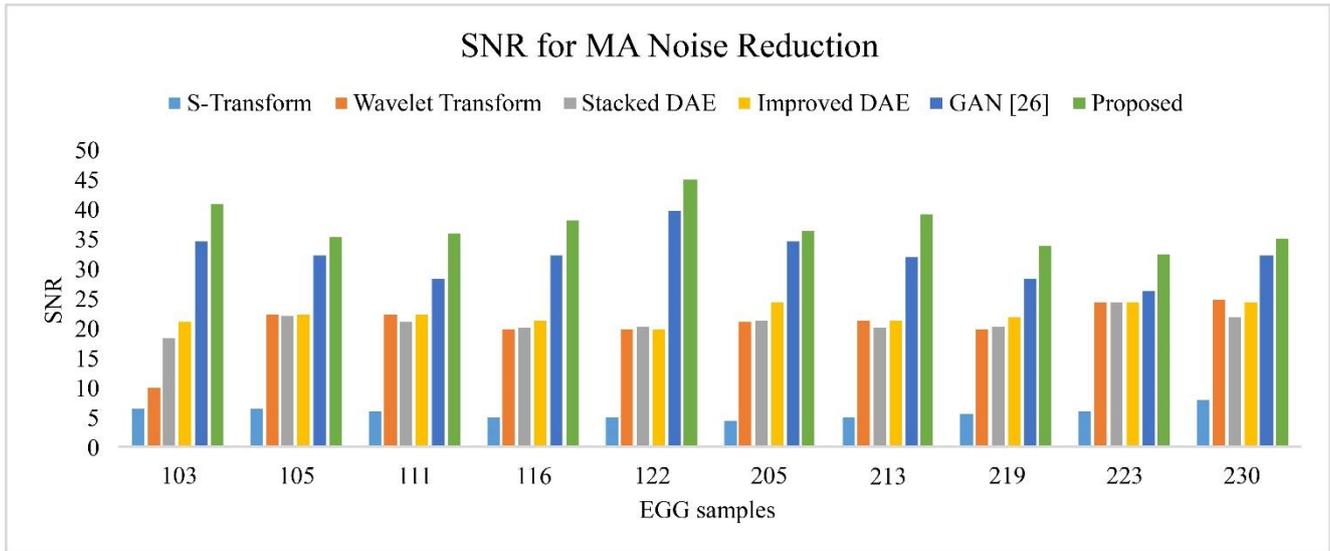

Fig. 11 SNR performance for EM noise reduction

Table 7. Comparative Study

| SNR | | 0 dB | | 1.25 dB | | 5 dB | |
|---|---|---|---|---|---|---|---|
| | | SNR | RMSE | SNR | RMSE | SNR | RMSE |
| EM | GAN | 28.42 | 0.176 | 32.86 | 0.012 | 60.01 | 0.010 |
| | PS | 32.50 | 0.112 | 36.81 | 0.10 | 66.25 | 0.008 |
| BW | GAN | 30.87 | 0.0058 | 33.74 | 0.0101 | 66.960 | 0.0046 |
| | PS | 36.21 | 0.0045 | 36.68 | 0.0100 | 69.230 | 0.0042 |
| MA | GAN | 36.44 | 0.0087 | 31.167 | 0.0157 | 57.680 | 0.0065 |
| | PS | 39.58 | 0.0075 | 36.221 | 0.0112 | 63.205 | 0.0050 |
| EM+BW | GAN | 28.36 | 0.0132 | 31.73 | 0.0112 | 58.44 | 0.0058 |
| | PS | 35.20 | 0.0104 | 39.40 | 0.010 | 65.201 | 0.0421 |
| EM+MA | GAN | 28.34 | 0.0037 | 31.91 | 0.0089 | 59.548 | 0.0068 |
| | PS | 32.50 | 0.0030 | 36.50 | 0.0075 | 66.30 | 0.0056 |
| BW+MA | GAN | 28.62 | 0.0132 | 32.30 | 0.0151 | 62.50 | 0.0088 |
| | PS | 36.50 | 0.0102 | 39.50 | 0.112 | 69.20 | 0.0071 |
| EM+BW+MA | GAN | 27.86 | 0.0128 | 31.77 | 0.0145 | 58.84 | 0.0076 |
| | PS | 34.20 | 0.0115 | 36.20 | 0.0101 | 63.20 | 0.0070 |

For this experiment, the average performance is obtained as 5.81dB, 20.72dB, 21.12dB, 22.48dB, 32.33dB, and 37.47dB by using S-Transform, Wavelet Transform, Stacked DAE, Improved DAE, GAN [26], and Proposed approach, respectively.

Based on these experiments, the performance of the proposed approach is measured and compared the obtained performance with existing schemes as mentioned in [25], where several existing techniques such as EKF, DWT, EMD, DWT+ADTF, EMD+ASMF, and SSCF are evaluated. Below given tables, 1,2, and 3 illustrate the comparative analysis in terms of MSE and PRD parameters for varied ECG signals, which are contaminated by adding different

*4.3.4. Average Noise reduction performance for the mixed type of noise*

In this experiment, the performance is measured in terms of average SNR for different types of noise combinations, which are added with a different SNR level as 0dB, 1.25 dB and 5 dB. The performance is measured in terms of SNR and RMSE. Below given table 7 shows the comparative analysis for this experiment.

This investigation considers diverse noise combinations and the performance is for these noises. The comparative study shows that the proposed ASCNet achieves better outcomes when compared with the existing schemes. The maximum SNR value is obtained as 69.23 by using the proposed approach, whereas for this sample, the existing approach achieves an SNR of 66.96dB.

## 5. Conclusion

This work has studied the significance of ECG signals in biomedical applications such as cardiomyopathy detection, heart rate monitoring etc. These signals are recorded with the help of hardware placed on the person's body. During the signal acquisition, some noise generated by these sensors is added to the original signal, degrading the signal quality. Thus, minimising noise becomes an important aspect of improving the diagnosis. Therefore, this work presents a deep learning-based methodology for ECG signal denoising. The proposed ASCNet approach uses modified ReLU, skip connection and attention module in DAE-based encoder and decoder architecture. The performance of the proposed approach was measured on the publicly available dataset and compared with existing techniques. The comparative analysis shows a significant improvement in the performance of the proposed approach.